\documentclass[aps,twocolumn,superscriptaddress,amsmath,amssymb,longbibliography]{revtex4-1}
\usepackage{graphicx}
\usepackage{dcolumn}
\usepackage{bm}
\usepackage{color}
\begin{document}

\title{High-energy spin waves\\in the spin-1 square-lattice antiferromagnet La$_2$NiO$_4$}

\author{A.\,N. Petsch}
\email{apetsch@stanford.edu}
\altaffiliation{Stanford Institute for Materials and Energy Sciences, Stanford University, Stanford, CA 94305, USA}
\altaffiliation{SLAC National Accelerator Laboratory, Menlo Park, CA 94025, USA}
\affiliation{H. H. Wills Physics Laboratory, University of Bristol, Bristol BS8 1TL, United Kingdom}

\author{N.\,S. Headings}
\affiliation{H. H. Wills Physics Laboratory, University of Bristol, Bristol BS8 1TL, United Kingdom}

\author{D. Prabhakaran}
\affiliation{Department of Physics, University of Oxford, Clarendon Laboratory, Oxford OX1 3PU, United Kingdom}

\author{A.\,I. Kolesnikov}
\affiliation{Neutron Scattering Division, Oak Ridge National Laboratory, Oak Ridge, Tennessee 37831, USA}

\author{C.\,D. Frost}
\affiliation{ISIS Facility, Rutherford Appleton Laboratory, Chilton, Didcot OX11 0QX, United Kingdom}

\author{A.\,T. Boothroyd}
\affiliation{Department of Physics, University of Oxford, Clarendon Laboratory, Oxford OX1 3PU, United Kingdom}
\author{R. Coldea}
\affiliation{Department of Physics, University of Oxford, Clarendon Laboratory, Oxford OX1 3PU, United Kingdom}
\author{S.\,M. Hayden}
 \email{s.hayden@bristol.ac.uk}
\affiliation{H. H. Wills Physics Laboratory, University of Bristol, Bristol BS8 1TL, United Kingdom}

\date{\today}

\begin{abstract}
Inelastic neutron scattering is used to study the magnetic excitations of the $S=1$ square-lattice antiferromagnet La$_2$NiO$_4$. We find that the spin waves cannot be described by a simple classical (harmonic) Heisenberg model with only nearest-neighbor interactions.  The  spin-wave dispersion measured along the antiferromagnetic Brillouin-zone boundary shows a minimum energy at the $(1/2,0)$ position as is observed in some $S=1/2$ square-lattice antiferromagnets.  Thus, our results suggest that the quantum dispersion renormalization effects or longer-range exchange interactions observed in cuprates and other $S=1/2$ square-lattice antiferromagnets are also present in La$_2$NiO$_4$. We also find that the overall intensity of the spin-wave excitations is suppressed relative to linear spin-wave theory indicating that covalency is important. Two-magnon scattering is also observed.  
\end{abstract}

\maketitle

\section{Introduction}
\label{sec:intro}
Studies of quantum (low-spin) square-lattice antiferromagnets (SLAFMs) are motivated by the desire to understand the ground state and excitations of a model Heisenberg system, and because superconductivity can develop by doping $S=1/2$ systems with antiferromagnetic interactions such as cuprates \cite{Bednorz1986_BeMu} and nickelates \cite{Li2019_LLW}. Large-$S$ antiferromagnets (AFM), like Rb$_2$MnF$_4$~\citep{Huberman2005_HCCT} ($S=5/2$), are generally well described by the semi-classical, harmonic, linear spin-wave theory (LSWT). In contrast, significant deviations from LSWT predictions have been observed in the spin excitations of $S=1/2$ SLAFMs, such as La$_2$CuO$_4$ (LCO)~\citep{Coldea2001_CHAP,Headings2010_HHCP} and Copper Deuteroformate Tetradeuterate (CFTD)~\citep{Roennow2001_RMCH,DallaPiazza2014_DMCN,Christensen2007_CRMH}. These systems show an anomaly in the excitations at the (1/2,0) position, on the antiferromagnetic Brillouin zone (mBZ) boundary. The anomaly is characterized by a strongly suppressed one-magnon energy and spectral weight as well as a broadening of the response in energy ($\hslash\omega$)~\citep{Roennow2001_RMCH,Headings2010_HHCP,DallaPiazza2014_DMCN,Christensen2007_CRMH,Tsyrulin2009_TPSX,Tsyrulin2010_TXSL,Wang2012_WLFE}.

LCO~\cite{Coldea2001_CHAP,Headings2010_HHCP} is a well characterized $S=1/2$ SLAFM based on transition-metal-oxide layers. It shows an unusual spin-wave dispersion which can be described with \textit{ferromagnetic} longer-range exchange interactions (2nd nearest-neighbor (NN) interaction or cyclic exchange) rather than a renormalization of the dispersion by effects beyond the linear spin-wave approximation. In the  Hubbard model, the longer-range exchange interactions result from the large $t/U$-ratio~\citep{Coldea2001_CHAP}. 
Here we study the $S=1$ system La$_2$NiO$_4$ (LNO) with smaller $t/U$-ratio.  This is a square-lattice $3d$ transition-metal-oxide antiferromagnet. Our aim is to determine whether the longer-range exchange interactions and (1/2,0) or (0,1/2) anomaly, respectively, observed in $S=1/2$ systems, persist in other systems.

LNO shows 3D magnetic order below $T_N \approx 320$~K with moderate spin-lattice coupling (the magnetic structure is discussed in Sec.~\ref{sec:mag_structure}). It is considered to be a Hubbard-Mott insulator in the Zaanen-Sawatzky-Allen scheme~\citep{Zaanen1985_ZaSA,Khomskii2015_Khom}. Above 75~K, LNO has the same $Bmab$ `low-temperature orthorhombic' (LTO) structure as LCO. The magnitude of the ordered moments in La$_2$NiO$_4$ has been found to be reduced with respect to the $S=1$ value. The moment reduction is believed to be due to a combination of covalency effects, arising from the anti-bonding orbitals of the Ni-O-Ni bonds~\citep{Wang1991_WSJL,Wang1992_WSJL,Lander1989_LBSH}, and zero-point spin fluctuations~\cite{Singh1989_Sing}.

\begin{figure}[ht]
  \centering
    \includegraphics[width=0.7\linewidth]{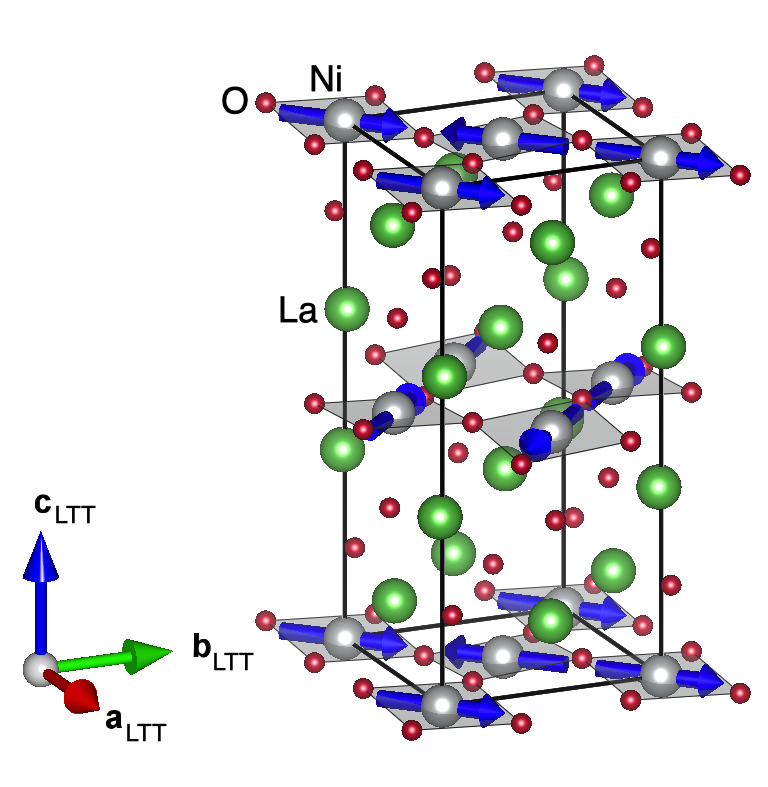}
    \caption{The low-temperature crystal and magnetic structure of La$_2$NiO$_4$ based on Ref.~\cite{Rodriguez-Carvajal_1991}. Blue arrows denote the Ni spins. The unit cell in this figure is labeled with the $P4_2/ncm$ (LTT) space group with $a_{\mathrm{LTT}} = b_{\mathrm{LTT}} \approx 5.5$~\AA\ and $c\approx 12.55$~\AA. Shaded squares indicated the four planar-oxygen sites surrounding each Ni$^{2+}$ ion and the buckling of the plane.}
    \label{fig:La2NiO4_structure}
\end{figure}

Previous inelastic neutron scattering 
(INS)~\cite{Aeppli1988_AeBu, Nakajima1993_NYHO} and resonant inelastic x-ray scattering (RIXS) \cite{Fabbris2017_FMX} studies show the existence of spin waves up to $\sim120$~meV.
A study \cite{Nakajima1993_NYHO} of the spin-wave dispersion in the $(H,H,L)$-plane observed two distinct gapped modes corresponding to fluctuations in and out of the $ab$-plane. The gaps were assigned to single-ion anisotropy. The Heisenberg NN interaction was determined to be $J\approx30$~meV. No out-of-plane, $\mathbf{c}$-axis, dispersion was observed implying $J_\perp/J<10^{-3}$ and making the magnetic excitations quasi-2D.

In this paper we present time-of-flight (ToF) INS data collected throughout the entire Brillouin zone and up to energy transfers of $\hslash\omega \approx 170$~meV on a high-quality single crystal of LNO. This enables us to resolve an anomalous high-$\hslash\omega$ spin-wave dispersion which resembles behavior observed in the $S=1/2$ SLAFM cuprate CFTD where it is assigned to quantum-dispersion-renormalization effects beyond linear-order spin-wave theory~\citep{Roennow2001_RMCH,DallaPiazza2014_DMCN,Christensen2007_CRMH}.
In addition, we show that the spectral weights are well described by a LSWT+$1/S$ model if anisotropy, covalency effects and two-magnon excitations are considered.

\section{Experimental details}
A LNO single crystal with a mass of $21.1$\,g was grown by the floating-zone technique and annealed at $1173$\,K in $5$\% CO and $95$\% CO$_2$ atmosphere to obtain the correct oxygen composition. A SQUID magnetometry measurement at 1\,T shows a N\'eel temperature of $T_N(1\,\mathrm{T})\approx320$\,K and a structural and spin reorientation transition at 75\,K. Thus, the oxygen excess $\delta$ in La$_2$NiO$_{4+\delta}$ is $\delta<0.007$~\citep{Rodriguez-Carvajal_1991,Buttrey1986_BuHR}. 

The ToF INS experiments were performed at the MAPS instrument at the ISIS Neutron and Muon Source at the Rutherford Appleton Laboratory~\citep{Ewings2019_ESPB,Boothroyd2009_BoHa} and the SEQUOIA instrument at the Spallation Neutron Source at the Oak Ridge National Laboratory~\citep{Granroth2010_GKSC}. Data were collected at $T=10$~K and $T=6$~K respectively.  The sample was aligned with $(1\overline{1}0)$ vertically. All presented MAPS data are integrated over $L\in[-15,15]$\,r.l.u. and all presented SEQUOIA data are integrated over $L\in[-10,10]$\,r.l.u.  

\subsection{Crystallographic Notation}
\label{Sec:notation}
The low-temperature structure of LNO is the LTT $P4_2/ncm$ structure~\cite{Rodriguez-Carvajal_1991}. This can be approximately described by the high-temperature tetragonal (HTT) $I4/mmm$ space group. We use the HTT conventional unit cell with $a=b=a_{\mathrm{HTT}} \approx3.89$\,\AA\ and $c\approx12.55$\,\AA\ to describe wave vectors in reciprocal space as $\mathbf{q}=H\mathbf{a}^\star+K\mathbf{b}^\star+L\mathbf{c}^\star \equiv (H,K,L)$ for the presentation of our data. For data integrated over $L$ and the spin wave theory we abbreviate to a square-lattice 2D-notation $(H,K)$. For a square-lattice the points $(H,K)$ and $(K,H)$ are equivalent.

\subsection{Magnetic Structure}
\label{sec:mag_structure}
At low temperatures, the host lattice of the antiferromagnetism in near stoichiometric LNO is believed to be $P4_2/ncm$ or LTT. 
Samples with a similar composition to ours develop a ferromagnetic (FM) component (i.e. show canting of the ordered moments) and have anomalies in the intensity of the antiferromagnetic Bragg peak measured by neutron scattering  
on entering the LTT state at $T \approx 75$~K \cite{Yamada1992_YONH}. Rodriguez-Carvajal \textit{et al.} \cite{Rodriguez-Carvajal_1991} (Table~4) show that only a magnetic structure belonging to the $\Gamma_{3g}$ irreducible representation of the $P4_2/ncm$ space group is consistent with this. We therefore assume that there is spin reorientation on entering the $P4_2/ncm$ structure and the antiferromagnetic structure is described by this magnetic mode as shown in Fig.~\ref{fig:La2NiO4_structure}. Note that this magnetic structure cannot be distinguished using diffraction from the $\Gamma_{4g}$ representation of the $Bmab$ space group proposed by Ref.~\cite{Rodriguez-Carvajal_1991} if two domains, rotated by 90$^\circ$ around the $\mathbf{c}$-axis, of equal population are present. In the $P4_2/ncm$ space group, the local-point-group symmetry of the Ni$^{2+}$ ions is $2/m$ and the moments are contained in the local mirror plane and point along the square diagonals of the LTT crystal structure such that the moments in adjacent layers are orthogonal as shown in Fig.~\ref{fig:La2NiO4_structure}. In relation to the HTT structure moments in the basal (middle) layer point almost along the $\mathbf{a}$ $(\mathbf{b})$ direction in HTT notation.

\section{Results}
\label{Sec:Results}
\begin{figure}[th] 
  \centering
    \includegraphics[width=0.99\linewidth]{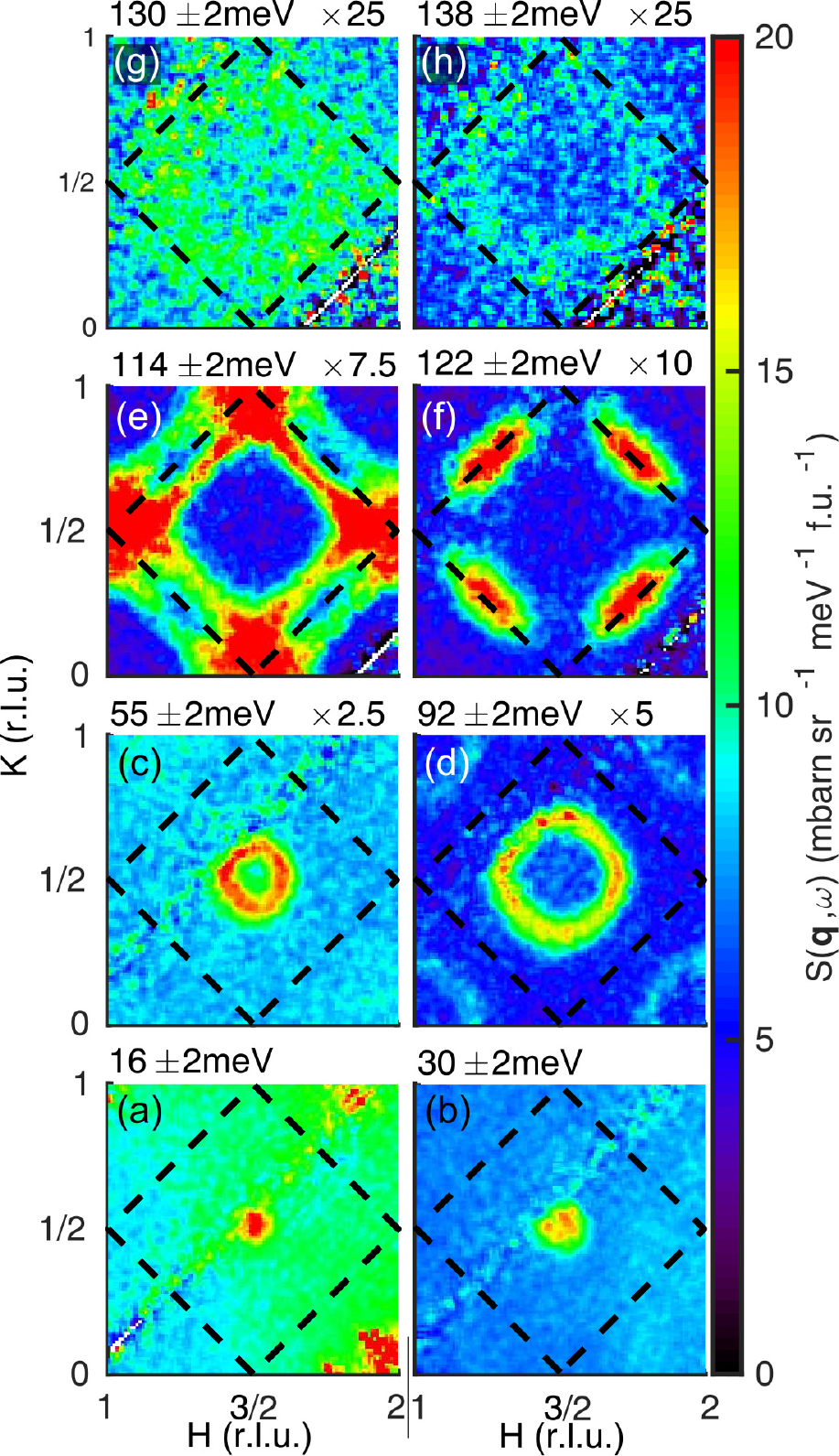}
    \caption{\label{fig:SlicesLNO} Representative constant-$\hbar\omega$ slices through the LNO data collected on SEQUOIA with $E_i=190$\,meV. The circles and other structures in panels up to $\hbar\omega=122$~meV are mostly due to single-magnon scattering. For $\hbar\omega\geq130$~meV two-magnon scattering is observed. The dashed lines are the mBZ boundaries.}
\end{figure}

The data are plotted and analyzed with the Horace package~\citep{Ewings2016_EBLD}. Fig.~\ref{fig:SlicesLNO} shows representative slices through the data collected with an incident energy $E_i=190$\,meV at the SEQUOIA instrument in terms of the scattering law $S(\mathbf{q},\omega)=\tfrac{k_i}{k_f}\tfrac{\mathrm{d}^2\sigma}{\mathrm{d}\Omega\mathrm{d}E^\prime}$. Data are normalized to absolute units via nuclear incoherent scattering from a vanadium standard, and are symmetrized about the $(H,H,0)$, $(H,\overline{H},0)$, and $(H,0,0)$ lines. The data collected at the MAPS instrument appear very similar. The slices in Fig.~\ref{fig:SlicesLNO}(a)-(d) show strong scattering as circles centered on (3/2,1/2), the center of an antiferromagnetic BZ. These are from one-magnon excitations, or spin waves. The scattering is consistent with spin gaps observed previously~\citep{Nakajima1993_NYHO}.
Around 92\,meV, spin-wave branches dispersing from reciprocal lattice points [e.g. (200)] become observable near the corners. In (e)-(f), lines of scattering parallel to the magnetic BZ (mBZ) boundaries (dashed lines) are the spin-wave branches originating from the (3/2,1/2) and (1,0)-type positions. The scattering in panels (g)-(h) is believed to be multi-magnon excitations. At higher $\hslash\omega$ the spin waves appear stronger first at the corners of the mBZ, and then at highest $\hslash\omega$ at the midpoints of the mBZ edges. This is unexpected for a classical NN Heisenberg SLAFM where no dispersion is expected along the mBZ boundaries. There, equal scattering, except for the magnetic form factor, is expected along the black dotted lines.

Our data are qualitatively consistent with a N\'eel SLAFM with single-ion anisotropy, significant multi-magnon scattering and an anomalous high-$\hslash\omega$ dispersion. For further analysis, a smooth function is fitted to a $\hbar\omega$-dependent cut at the \textit{ferromagnetic} reciprocal-lattice position (1,0) and subtracted from all analyzed data. This removes most incoherent and multi-phonon background.
\begin{figure}[ht]
  \centering
    \includegraphics[width=0.9\linewidth]{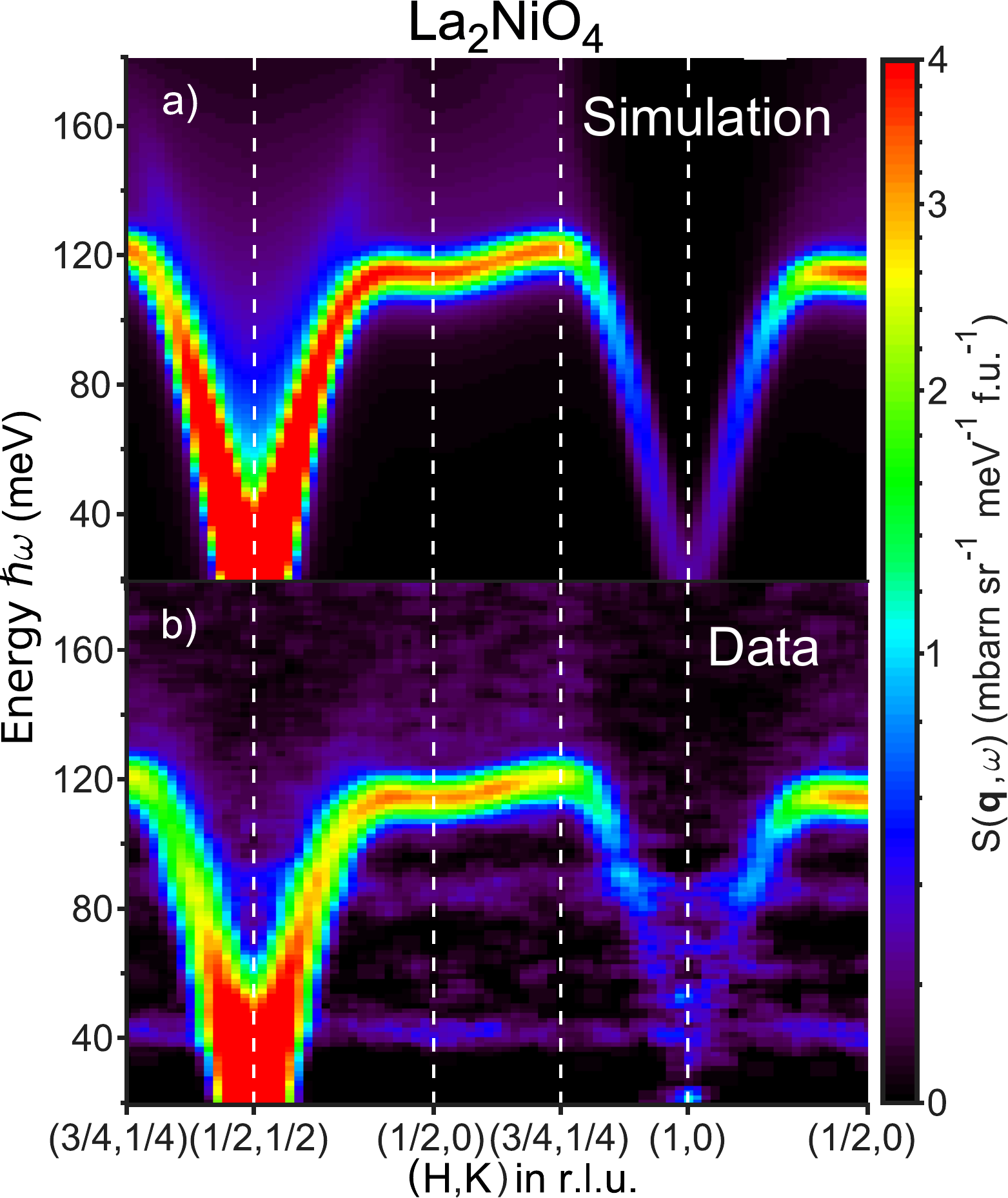}
    \caption{\label{fig:Comp} Magnetic excitations in La$_2$NiO$_4$. (a) Scattering function $S(\mathbf{q},\omega)$ simulated from Eqns.~\ref{eqn:S_bb}-\ref{eqn:S_aa} convoluted with finite lifetime and resolution. (b) INS data  collected with $E_i = 260$\,meV at SEQUOIA. A background consisting of a smooth interpolation of a $\mathbf{q} =(1,0)$ spectrum integrated over $L\in[-15,15]$ has been subtracted from each $\mathbf{q}$. A non-linear color-coded intensity scale is used to enhance the weak two-magnon scattering.}
\end{figure}

The calculated and measured intensities of the magnetic excitations after background subtraction are shown in Fig.~\ref{fig:Comp}(a,b). The multi-magnon scattering and anomalous dispersion are clearly visible at the mBZ boundary. The two, mostly dispersionless, lines at $\sim 42$\,meV and $\sim 87$\,meV are optical phonon modes.

One-dimensional (1D) cuts are taken through the data along high-symmetry lines marked in the inset of Fig.~\ref{fig:Fits}(b) to fit the data with the model described in the following section. The data sets from both instruments are individually fitted to determine the model parameters. Some representative cuts with fits are shown in Fig.~\ref{fig:Fits}(d).

\section{Spin-wave model for a single N\lowercase{i}O$_2$ plane}
\label{sec:theory}
\begin{figure*}[t]
\centering
\includegraphics[width=0.98\linewidth]{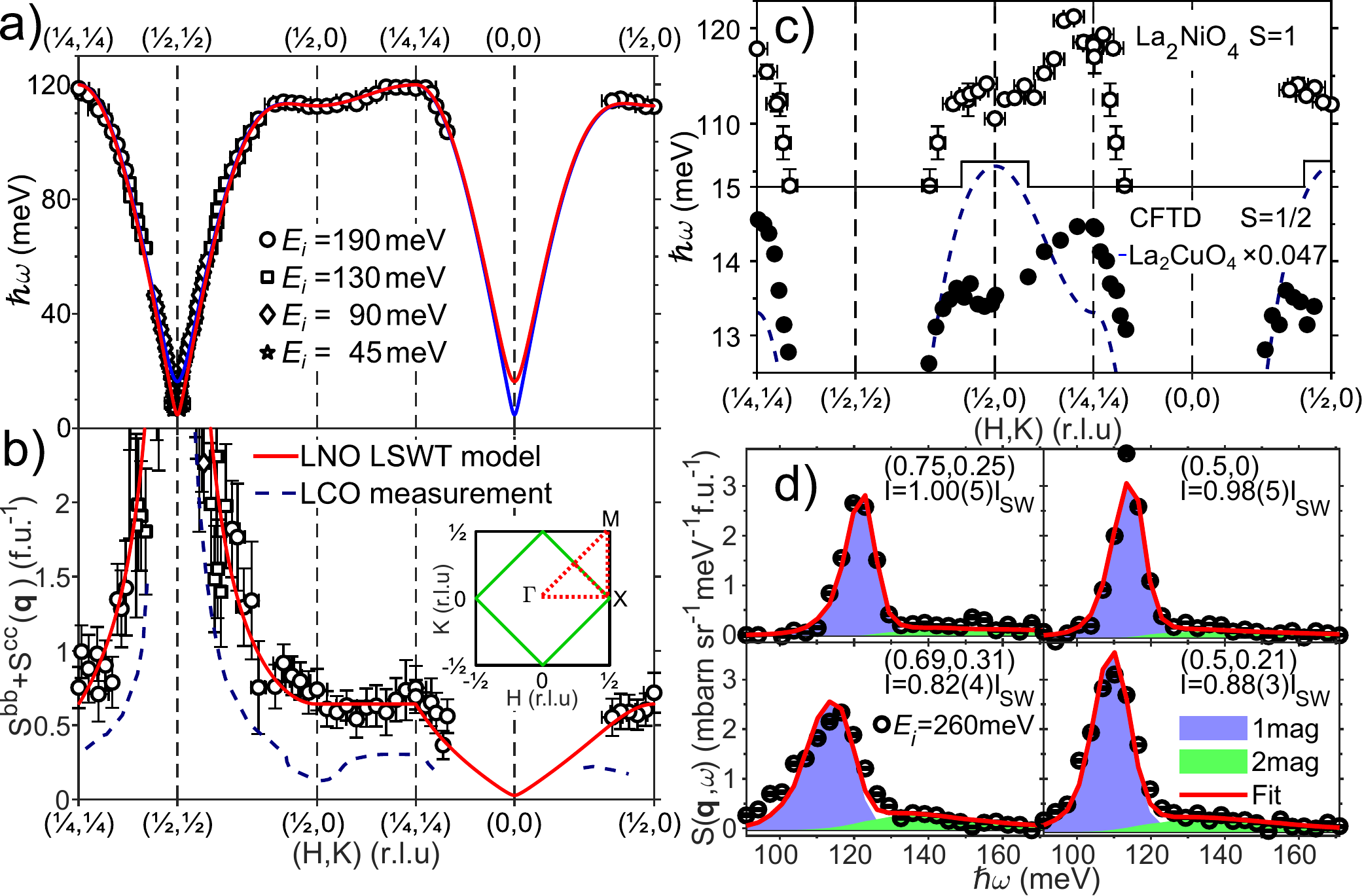}
    \caption{\label{fig:Fits}Results of the data fitting and comparison with similar compounds: (a) Spin wave dispersion \cite{supp_info} in red (blue) for the in-plane (out-of-plane) mode with the peak positions from the MAPS data; (b) Amplitude of the spin-wave pole in units of spin (Multiply by $g^2 \mu_B^2$ to get in units of $\mu^2_B$.) from the Horace~\citep{Ewings2016_EBLD} fitting of Eqns.~\ref{eqn:S_bb}-\ref{eqn:S_cc} to the MAPS data, LSWT prediction with $Z_{\textrm{co}}=0.78(6)$ (red) and absolute results from LCO with $Z_{\textrm{co}}=0.89(23)$~\citep{Headings2010_HHCP} for $Z_d(S=\tfrac{1}{2})=0.57$~\citep{Canali1993_CaWa,Canali1992_CaGW} (blue dashed); (c) MAPS data in comparison with CFTD~\citep{Roennow2001_RMCH} and LCO dispersion scaled by $J(\mathrm{CFTD})/Z_cJ(\mathrm{LCO})=0.047$~\citep{Headings2010_HHCP} (blue dashed); It is noted that $(0,\tfrac{1}{2})$ is equivalent to $(\tfrac{1}{2},0)$; (d) Representative cuts through the SEQUOIA data fitted with one+two-magnon model with $\hslash\Gamma=1.5$\,meV.}
\end{figure*}
LNO is a Hubbard-Mott insulator,  its magnetism is well described by an extended Heisenberg model with spin quantum number $S=1$ on the Ni$^{2+}$ (3d$^{8}$) sites with the orbital moments quenched by the octahedral-crystal-field environment of oxygen ions. 
We consider a single NiO$_2$ layer (the basal layer in Fig.~\ref{fig:La2NiO4_structure}) with the ordered moment along the $\mathbf{z}$-axis or $\mathbf{a}$ (See Sec.~\ref{Sec:notation}). The spin Hamiltonian can be written as~\citep{Roger1989_RoDe,Chubukov1992_ChGB, Nakajima1993_NYHO, Nolting2009_NoRa,Coldea2001_CHAP,Marshall1971_MaLo}
\begin{align}
\mathcal{H}&=\sum_{\langle i,j\rangle}J_{ij}\mathbf{S}_i\cdot\mathbf{S}_j\!+\!\sum_{i}\left[K_{c}\big(S^y_i\big)^2\!+\!K_{a}\big(S^z_i\big)^2\right]  \nonumber \\
& + J^{\Box} \! \! \sum_{\langle i,j,k,l \rangle} 
[(\mathbf{S}_i\cdot\mathbf{S}_j)(\mathbf{S}_k\cdot\mathbf{S}_l)+
(\mathbf{S}_i\cdot\mathbf{S}_l)(\mathbf{S}_k\cdot\mathbf{S}_j) \nonumber \\
&-(\mathbf{S}_i\cdot\mathbf{S}_k)(\mathbf{S}_j\cdot\mathbf{S}_l)],
\label{eqn:Ham} 
\end{align}
where $J_{ij}$ represents the first to third NN Heisenberg exchange interactions $J$, $J^\prime$ and $J^{\prime\prime}$, $K_{a}\leq0$ is an easy-axis anisotropy and $K_{c}\geq0$ an out-of-plane hard-axis anisotropy. Here $(x,y,z)$ are along the HTT $(b,c,a)$ axes respectively (see Sec.~\ref{Sec:notation}). The local-anisotropy terms are symmetry allowed in the $2/m$ local-point-group symmetry of the Ni$^{2+}$ ions in the LTT structure (the local mirror plane is $ac$) and are attributed to higher-order effects of the local crystal field and spin-orbit coupling. A bilinear-biquadratic interaction, suggested for $S=1$, is neglected as it becomes indistinguishable from other interaction terms in the N\'eel state~\citep{Oitmaa2013_OiHa,Papanicolaou1988_Papa,Toth2012_TLMP}. Also, an out-of-plane spin canting of 0.1$^\circ$, assigned to a finite Dzyaloshinskii–Moriya interaction (DMI), has been observed~\cite{Yamada1992_YONH}. Although DMI can induce non-degenerate spin-wave modes the spin canting is too small to describe the previously reported gap size~\cite{Nakajima1993_NYHO, Toth2015_ToLa,Yamada1992_YONH} (see Ref.~\cite{Nakajima1993_NYHO} and Appendix~\ref{sec:dmi}) and hence it is neglected in our analysis. The cyclic-term $J^\Box$ considered in LCO ~\citep{MacDonald1988_MaGY,MacDonald1990_MaGY,Klein1973_KlSe,Kato1949_Kato} is indistinguishable in LSWT from $J^\prime$ and is considered later (see also Appendix~\ref{sec:dev}).

The spin-wave excitations of the Hamiltonian are determined in the harmonic limit, commonly referred to as LSWT. For more details see Appendices~\ref{sec:dev} and~\ref{sec:alt_dev}. There are two distinct spin-wave modes for $K_c>0$ corresponding to spin fluctuations along $\mathbf{b}$ (in-plane) and $\mathbf{c}$ (out-of-plane) respectively. Their dispersion relations are given by $\hslash\omega_{\mathbf{q}}$ and $\hslash\omega_{\mathbf{q}}^{\prime}= \hslash\omega_{\mathbf{q}+\boldsymbol{\tau}_{\textrm{AF}}}$ respectively,
where
\begin{align}
\hslash\omega_{\mathbf{q}}&=Z_c\sqrt{A_{\mathbf{q}}^2-B_{\mathbf{q}}^2}, \label{eq:DispersionLNO} \\
A_{\mathbf{q}}&=4S\left[-\frac{K_a}{2}+\frac{K_c}{4}+J-J^{\prime}(1-\nu_h\nu_k)\right],  \\ 
B_{\mathbf{q}}&=4S\left[J\frac{\nu_h+\nu_k}{2}-\frac{K_c}{4}\right],
\end{align}
with $\nu_{\xi}\!=\!\cos(2\pi \xi)$ and $\boldsymbol{\tau}_{\textrm{AF}}=(1/2,1/2)$ the N\'{e}el-magnetic-structure propagation vector, expressed in reciprocal-lattice units of the HTT unit cell. $Z_c\approx1.09$ is a spin-fluctuation correction factor which renormalizes the excitation energy~\citep{Igarashi1992_Igar,Singh1989_Sing}. The Bogoliubov transformation parameters then yield the correlation (scattering) functions for the one- and two-magnon excitations in the $T \rightarrow 0$ limit \citep{White1965_WhSO,Heilmann1981_HKER,Ewings2008_EPBG,Lorenzana2005_LoSC,Coldea2003_CoTT} (See Appendix~\ref{sec:dev})
\begin{align}
S^{bb}(\mathbf{q},\omega)\!=&\frac{Z_dZ_{\textrm{co}}S}{2}\left|u_{\mathbf{q}}\!-\!v_{\mathbf{q}}\right|^2 \delta(\hslash\omega\!-\!\hslash\omega_{\mathbf{q}})  \label{eqn:S_bb} \\
S^{cc}(\mathbf{q},\omega)\!=&\frac{Z_dZ_{\textrm{co}}S}{2}\left|u_{\mathbf{q}\!+\!\boldsymbol{\tau}_{\textrm{AF}}}\!+\!v_{\mathbf{q}\!+\!\boldsymbol{\tau}_{\textrm{AF}}}\right|^2 \delta(\hslash\omega\!-\!\hslash\omega_{\mathbf{q}\!+\!\boldsymbol{\tau}_{\textrm{AF}}}) \label{eqn:S_cc}\\
S^{aa}(\mathbf{q},\omega)\!=&N Z_{\textrm{co}}^2(S\!-\!\Delta S)^2\delta(\hslash\omega)\delta(\mathbf{q}\!-\!\boldsymbol{\tau}_{\textrm{AF}}\!-\!\bm{\tau})+ \nonumber \\
 \frac{Z_{2\rm{M}}Z_{\textrm{co}}}{2N}&\sum_{\mathbf{q}_1,\mathbf{q}_2} f(\mathbf{q}_1,\mathbf{q}_2)\delta(\hslash\omega\!-\!\hslash\omega_{\mathbf{q}_1}\!-\!\hslash\omega_{\mathbf{q}_2}) \times \nonumber \\
&\delta(\mathbf{q}\!-\!\boldsymbol{\tau}_{\textrm{AF}}\!-\!\mathbf{q}_1\!-\!\mathbf{q}_2\!-\!\bm{\tau}), \label{eqn:S_aa}
\end{align} 
where $u_{\mathbf{q}}=\cosh \theta_{\mathbf{q}}$, $v_{\mathbf{q}}=\sinh\theta_{\mathbf{q}}$, $\tanh (2\theta_{\mathbf{q}})=B_{\mathbf{q}}/A_{\mathbf{q}}$,  $f(\mathbf{q}_1,\mathbf{q}_2)=\left|u_{\mathbf{q}_1}v_{\mathbf{q}_2} + u_{\mathbf{q}_2}v_{\mathbf{q}_1}\right|^2$. $N$ is the total number of spins in the lattice, $\bm{\tau}$ is a (HTT structural) reciprocal lattice vector, and $\Delta S=\langle v^2\rangle$ is the zero-point spin reduction, where $\langle ...\rangle$ means the average over the full Brillouin zone. The first term in $S^{aa}$ (denoting fluctuations along the $\mathbf{a}$ axis) contains the elastic magnetic Bragg peak and the second term is the inelastic two-magnon continuum, with one of the two wave vectors in the sum restricted to one full Brillouin zone. The above dynamical correlations and the dispersion relation $\omega_{\mathbf{q}}$ have the translational periodicity of the full Brillouin zone. 

The prefactor $Z_d$ is a one-magnon intensity renormalization factor due to higher order effects neglected at linear order in spin-wave theory, $Z_{2\rm{M}}$ is a corresponding factor for the two-magnon scattering. We have also included an additional factor $Z_{\textrm{co}}$ in Eqns.~\ref{eqn:S_bb}-\ref{eqn:S_aa} to take account of covalency effects, $Z_{\textrm{co}}=1$ in the absence of these. For $K_c=0$, $Z_d=1-\Delta S/S$, and assuming $Z_{2\rm{M}}=1$, the total spin sum rule is satisfied such that elastic, one-magnon and two-magnon scattering integrated over all energies and a full Brillouin zone add up to $S(S+1)$ per spin, shared between the three contributions as $(S-\Delta S)^2$, $\left(S-\Delta S\right)(2\Delta S+1)$ and $\Delta S (\Delta S+1)$, respectively. To derive the above we have used the fact that $u$ is even and $v$ is odd with respect to a wave-vector shift by the magnetic propagation vector $\boldsymbol{\tau}_{\textrm{AF}}$, so the average $\langle uv \rangle=0$. For finite $K_c$, this is no longer the case and to satisfy the total sum rule one needs to use 
\begin{align}
Z_d=1-\frac{\Delta S}{S}-\frac{\langle uv\rangle^2}{S(2\Delta S+1)} \label{eqn:Zd}
\end{align}
as the integrated two-magnon scattering becomes $\Delta S (\Delta S+1)+ \langle uv \rangle^2$. 

\begin{figure}[th] 
  \centering
    \includegraphics[width=0.99\linewidth]{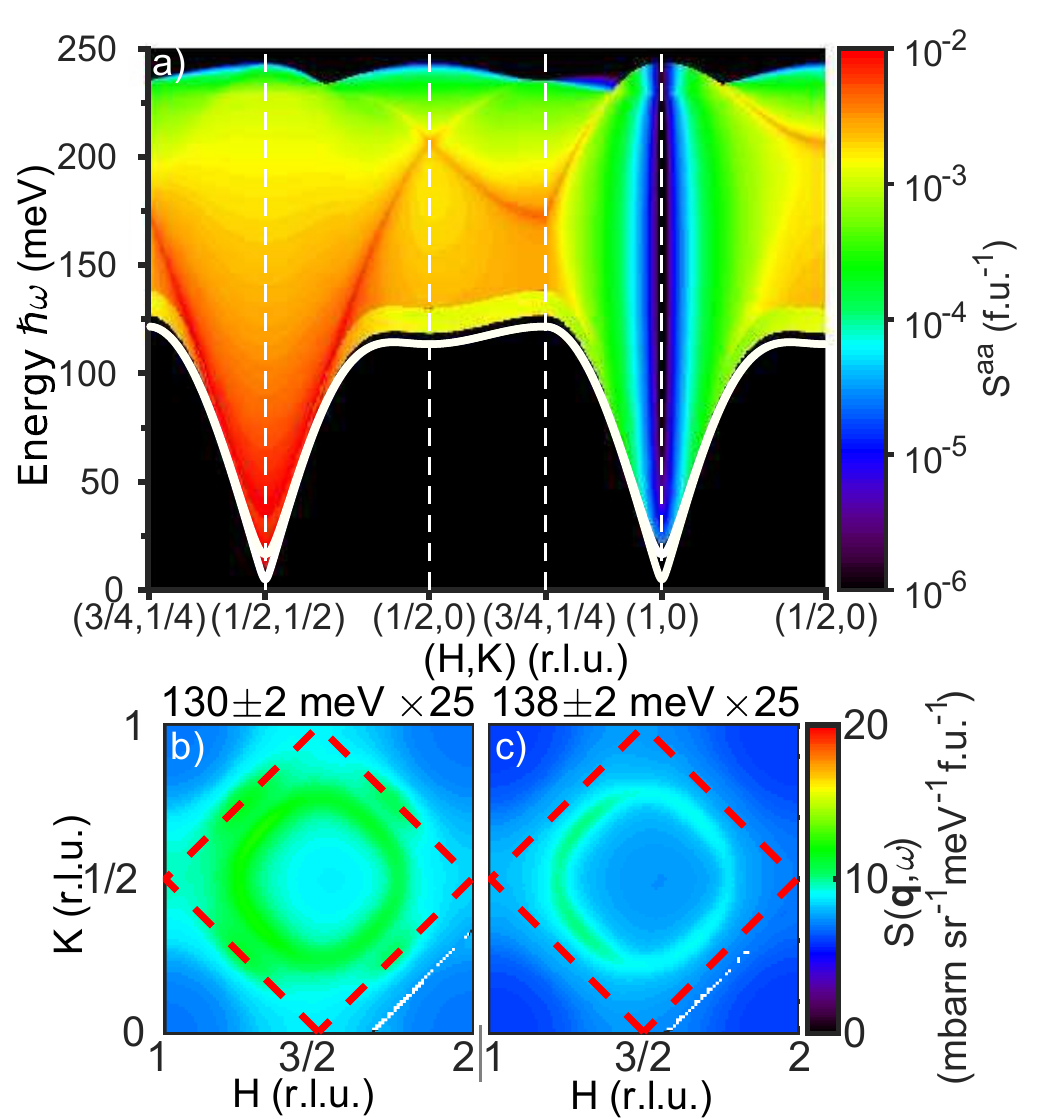}
    \caption{\label{fig:2mag_sim} a) Simulation of two-magnon scattering using Eqn.~\ref{eqn:S_aa}. The solid white lines show the magnon dispersions $\hslash\omega_{\mathbf{q}}$ and $\hslash\omega_{\mathbf{q}+\boldsymbol{\tau}_{\textrm{AF}}}$. b,c) Simulations of $S(\mathbf{q},\omega)$ due to two-magnon scattering with an added constant background of $0.28$ or $0.24$\,mb$^{-1}$sr$^{-1}$f.u.$^{-1}$, respectively. These closely resemble the slices in Fig.~\ref{fig:SlicesLNO}g) and h), respectively.}
\end{figure}

Significant two-magnon scattering is observed even in spin-5/2 systems. So we expect to see this also in the present $S=1$ system \cite{Huberman2005_HCCT} and signatures are presented in Fig.~\ref{fig:Fits}d) and~\ref{fig:SlicesLNO}g) and h) as shown in Fig.~\ref{fig:2mag_sim}b) and c), respectively. 
Eqn.~\ref{eqn:S_aa} is evaluated on a three-dimensional grid ($\mathbf{q}_{2D},\hslash\omega)$ and then convolved with a $\mathbf{q}$-independent inverse lifetime of $\hslash\Gamma=$1.5\,meV per excited magnon. To fit the data Eqns.~\ref{eqn:S_bb}-\ref{eqn:S_cc} are added after lifetime broadening to the three-dimensional grid and these spin-spin correlation functions are then multiplied by the anisotropic magnetic form factor of the Ni$^{2+}$ $e_g$-orbitals~\citep{Weiss1959_WeFr,Clementi1974_ClRo,Desclaux1978_DeFr,Freeman1979_FrDe,Anderson2006_ABCL}. Deviations due to covalency are included through the factor $Z_{\mathrm{co}}$. Finally, these functions are convolved  with the instrument resolution function by Tobyfit in the Horace package~\citep{Ewings2016_EBLD}.
Simulations of $S^{aa}(\mathbf{q},\omega)$ without instrumental and lifetime broadening are shown in Fig.~\ref{fig:2mag_sim}. The two-magnon term appears unusual due to the anisotropy gaps resulting in a `peak' above $\hslash\omega_\mathbf{q}$ rather than a tail arising at $\hslash\omega_\mathbf{q}$. It is further observed that the instrument resolution dominates the broadening.

\section{Dispersion}
In order to plot the dispersion, we fitted Gaussian functions to 1D cuts through the data to obtain the peak positions, plotted in Fig.~\ref{fig:Fits}(a). Constant-$\mathbf{q}$ cuts with $E_i=45$\,meV at low-$\hslash\omega$ clearly show two distinct gapped spin-wave modes (not shown). The high-$\hslash\omega$ excitations show dispersion along the mBZ boundary. From the lower-$\hslash\omega$ data ($\le 50$~meV) we can obtain an estimate of the $J\approx29$\,meV and the spin gaps at (1/2,1/2) $\Delta_{1}\!\approx\!4Z_c\sqrt{\!J(-K_a)}\approx5$\,meV and $\Delta_{2}\approx\!4Z_c\sqrt{J(K_c\!-\!K_a)}\approx16$\,meV are in good agreement with previous work~\citep{Nakajima1993_NYHO}. We mostly fixed $K_a$ (see Table~\ref{tab:LNOFit}) when fitting the whole dispersion curve below. 

The high-$\hslash\omega$ dispersion cannot be described by $J$ only in LSTW but requires a finite $J^\prime$. The upturn from $\mathbf{q}=(1/2,0)$ to $\mathbf{q}=(1/4,1/4)$ can be described by an antiferromagnetic $J^\prime>0$ with $J^\prime\approx5.8(3)\%$ of $J$ which yields, including the anisotropy, a difference of $\sim6\%$ between $(1/2,0)$ and $(1/4,1/4)$. The $J^\prime>0$ describes the dispersion along the mBZ boundaries as well as the local dispersion minimum at $(1/2,0)$. Adding $J^{\prime\prime}$, described in Eqn.~\ref{AB}, does not improve the fitting, and is hence neglected from now on. 

\begin{table}[ht!]
\centering
 \begin{tabular}{|c | c | c | c | c | c|}
 \hline 
 $J$ & $J^{\prime}$ & $J^{\prime\prime}$ & $J^\Box$ & $K_{a}$ & $K_c$ \\
 (meV)  & (meV) & (meV) & (meV) & (meV) & (meV) \\
 \hline
 29.02(8) & 1.68(5) & 0 (-) & 0 (-) & -0.035(3) & 0.443(11)  \\
 28.2(11) & 1.3(5) & -0.2(3)  & 0 (-) & -0.04 (-) & 0.46(2)  \\ 
 29.00(8) & 1.67(5) & 0 (-) &  0 (-) & -0.04 (-) & 0.445(11)  \\
 32.34(13) & 0 (-) & 0 (-) & -1.67(5) & -0.04(-) & 0.445(11)  \\
\hline
 \end{tabular}
\caption{\label{tab:LNOFit} Fitted Heisenberg coupling and anisotropy constants which determine the dispersion given in Eqn.~\ref{eq:DispersionLNO}. The rows describe different fitting models, the first row includes no $J^{\prime\prime}$, the second row includes all terms but with $K_a$ fixed and adapted from Ref.~\citep{Nakajima1993_NYHO} and the third row excludes $J^{\prime\prime}$ and again uses the value from  Ref.~\citep{Nakajima1993_NYHO}, for $K_a$. The forth row parameterizes the second row in terms of $J$ and $J^{\Box}$. Parameters marked $(-)$ are fixed during fitting.}
\end{table}

La$_2$CuO$_4$ shows an inverted dispersion (with respect to LNO) along the mBZ boundary $[(0,1/2) - (1/4,1/4)]$ and $J^\prime<0$ (when $J^\Box=0$)~\citep{Coldea2001_CHAP} (See Fig.~\ref{fig:Fits}(c)). The dispersion in LCO can be explained by higher-order terms in the $t/U$ expansion of the Hubbard model for hopping between Cu sites. This yields, terms in $J^\Box$, $J^{\prime}$ and $J^{\prime\prime}$~\citep{MacDonald1988_MaGY,MacDonald1990_MaGY,Klein1973_KlSe,Takahashi1977_Taka}. The term in $J^\Box$ has the same effect on the dispersion as $J^{\prime}$ in LSWT (see Eqn.~\ref{AB}) so cannot be distinguished from $J^{\prime}$. 

The $S=3/2$ $3d$ square-lattice transition-metal-oxide AFM La$_2$CoO$_4$ (LCoO) has the same LTT structure as La$_2$NiO$_4$ and neutron scattering measurements of the magnon dispersion have also been carried out on this material \cite{Babkevich2010_BPFB}. The spin-wave excitations in LCoO also show dispersion along the mBZ with a minimum at ($1/2, 0)$, as in LNO, but the effect is weaker in LCoO than in LNO. In Table~\ref{tab:J_comparison} we compare the exchange couplings of LCO, LNO and LCoO. Both LNO and LCoO have $J^{\prime} >0$. Thus, the anomalous compound is LCO, which yields $J^{\prime}<0$ when fitted with $J^{\Box}=0$. The natural conclusion is that cyclic exchange is present in LCO, but negligible in LNO and LCoO which is consistent with the more substantial $t/U$ in LCO.

\begin{table}
\begin{tabular}{|l|c|c|c|c|c|}
 \hline 
  & $S$ & $J$ &  $J^{\prime}$ & $J^{\prime\prime}$ & $J^{\Box}$\\
  & & (meV) & (meV) & (meV) & (meV) \\
 \hline
La$_2$CuO$_4$ \cite{Headings2010_HHCP} & 1/2 & 114(2) & -12(2) & 2.9 (2) &  0(-) \\
La$_2$CuO$_4$, $t/U$ \cite{Headings2010_HHCP} & 1/2 & 143(2) & 2.9 (2) & 2.9 (2) &  58(4)\\
La$_2$NiO$_4$                           & 1 & 28.2(11) & 1.3(5) & -0.2(3) & 0(-) \\
La$_2$CoO$_4$ \cite{Babkevich2010_BPFB} & 3/2 & 9.69(2) &  0.43(1) & 0.12(2)  & 0(-) \\ 
\hline
 \end{tabular}
\caption{Comparison of exchange constants for square-lattice $3d$ transition-metal-oxide AFMs. Here $S$ is the spin and $t/U$ indicates results from fitting with the extended $t/U$-expansion~\citep{MacDonald1988_MaGY,MacDonald1990_MaGY,Klein1973_KlSe,Takahashi1977_Taka}.}
\label{tab:J_comparison}
\end{table}

As shown in Fig.~\ref{fig:Fits}(c) the high-$\hslash\omega$ dispersion in LNO however closely resembles the dispersion in the $S=1/2$ SLAFM copper deuteroformate tetradeuterate (CFTD)~\citep{Roennow2001_RMCH,DallaPiazza2014_DMCN,Christensen2007_CRMH} as well as similar compounds such as Cu(pyrazine)$_2$(ClO$_4$)$_2$~\citep{Tsyrulin2009_TPSX,Tsyrulin2010_TXSL}, and CuF$_2$(H$_2$O)$_2$(pyrazine)~\citep{Wang2012_WLFE}. In these compounds the dispersion along the mBZ is explained by quantum effects which renormalize the dispersion. The dispersion in $S=$1/2 SLAFMs is predicted by various theoretical calculations. Although all models predict a dispersion on the mBZ boundary they disagree over the magnitude of the dispersion and its origin. To our knowledge no calculations are available at the time of the submission for $S=$1 SLAFMs but from the Holstein-Primakoff transformation a suppression of quantum effects by at least a factor of two from $S=$1/2 to $S=$1 systems is predicted~\citep{Verresen2018_VePM}.

\section{Spectral weight}
The 1D cuts, taken from the MAPS instrument data, are used to fit the overall spectral weights due to the better resolution at low $\hslash\omega$. The cuts are fitted with the intensities calculated from Eqns.~\ref{eqn:S_bb}-\ref{eqn:S_cc} convolved with a $\mathbf{q}$-independent inverse lifetime $\Gamma$. A linear function is added to remove multi-magnon scattering and a $\mathbf{q}$-dependent background. 
The fitted one-magnon spectral weights throughout the lattice BZ are depicted in Fig.~\ref{fig:Fits}(b). The $\mathbf{q}$-dependence of the one-magnon spectral weights is qualitatively well described by the LSWT model. Small deviations arise near the antiferromagnetic BZ center. Firstly, because the anisotropy prevents the divergence of the spectral weight there, and secondly, the fitting is hindered by contamination from the magnetic Bragg peak and the multi-magnon scattering. Utilizing the Hamiltonian parameters determined from the dispersion relations we find $\Delta S=0.188(1)$ and $\langle uv \rangle=-0.017(1)$, yielding $Z_d\approx 0.812(1)$. A fit of the high-$\hslash\omega$ one-magnon excitations using Eqns.~\ref{eqn:S_bb}-\ref{eqn:S_aa}, with $Z_d\approx 0.812(1)$, the SEQUOIA data yields $Z_{\textrm{co}}=0.78(6)$. The $Z_d$ follows from $\Delta S=0.188(1)$ and $\langle uv \rangle=-0.017(1)$, therein derived from the Hamiltonian parameters of the fitted dispersion relation. We believe $Z_{\textrm{co}}<1$ because there is a reduction of the ordered Ni$^{2+}$ moment due to oxygen covalency effects as observed in neutron diffraction \cite{Wang1991_WSJL,Wang1992_WSJL,Lander1989_LBSH}. 

In the related spin-1/2 compounds LCO and CFTD anomalous scattering is observed near (1/2,0) and equivalent (0,1/2) point. To study if such scattering also arises in LNO the one+two-magnon model is fitted to 59 constant-$\mathbf{q}$ cuts through the high-$\hslash\omega$ excitations in the SEQUOIA data. The model includes $Z_{\mathrm{co}}$ and the subsequent analysis focuses particularly on the comparison of (1/4,1/4) and the anomalous (1/2,0) point. All cuts are fitted with a small individual constant background term to account for $\mathbf{q}$-dependent variations in the background. Some representative fits are shown in Fig.~\ref{fig:Fits}(d) where $\Gamma=1.5$\,meV and the fits are unchanged for smaller $\Gamma$ values.

As shown in Fig.~\ref{fig:Fits}(d), the model gives a good description of peak shapes and continua and, further, of the relative intensities of one- and two-magnon spectral weights. The calculated spectral weights also agree quantitatively well with the measured spectral weights after re-scaling for $Z_{\mathrm{co}}$. Moreover, the model including $Z_{\mathrm{co}}$ gives a good description along all high-symmetry directions as shown in Fig.~\ref{fig:Comp}(a) and (b). 

In contrast to the $S=1/2$ compounds CFTD~\citep{Roennow2001_RMCH,DallaPiazza2014_DMCN,Christensen2007_CRMH} and LCO~\citep{Headings2010_HHCP}, in LNO neither the reduced one-magnon spectral weight (see Fig.~\ref{fig:Fits}(b) nor the enhanced multi-magnon spectral weight is observed at (1/2,0) within the statistical limitations. 

\section{Discussion}
In the preceding sections we have seen that although aspects of the magnetic excitations of LNO are qualitatively described by a simple linear-spin-wave theory with two-magnon scattering, the intensity and dispersion show some significant deviations. The measured spin-wave dispersion is well described by an anisotropic semi-classical NN Heisenberg AFM below $\sim$100~meV. This theory also qualitatively describes the two-magnon excitations observed for $\hslash\omega \approx 130$~meV. Two aspects of the excitations that are not well described by LSWT are the overall intensity of the excitations and the dispersion of the high-$\hslash\omega$ excitations.

The overall intensity of the excitations is determined by the absolute normalization of the measured signal and yields a scaling factor of $Z_{\textrm{co}} =0.78(6)$ through Eqns.~\ref{eqn:S_bb}-\ref{eqn:S_cc} when the quantum renormalization factor $Z_d$ is taken into account.  The most likely explanation for the difference from $Z_{\textrm{co}}=1$ are the covalency effects \cite{Wang1991_WSJL,Wang1992_WSJL,Lander1989_LBSH} present in nickel oxides. This results in some of the ordered and fluctuating magnetic moments residing on the oxygen atoms, reducing the signal seen in the present experiment.

We also observe a deviation from the predictions of LSWT for a NN SLAFM in the form of significant dispersion along the mBZ boundaries. The dispersion indicates the presence of longer-ranged exchange interactions. The structurally related compound LCO also shows dispersion along the mBZ boundaries but with the opposite sense \citep{Coldea2001_CHAP}. In the case of LCO, the dispersion is due to the substantial $t/U \approx 0.11$ resulting in a substantial \textit{ferromagnetic} next-NN exchange $J^\prime <0$ in superexchange theory \citep{MacDonald1988_MaGY,MacDonald1990_MaGY,Klein1973_KlSe,Takahashi1977_Taka}. Such a mechanism cannot explain the \textit{antiferromagnetic} $J^\prime >0$ observed in LNO and LCoO. An anisotropy in $J$ as origin of the high-$\hslash\omega$ dispersion is also excluded as this would imply two distinct high-$\hslash\omega$ spin-wave modes with differing dispersion, as shown in Ref.~\citep{Koshibae1994_KoOM}.

Thus, our observation of a downward dispersion from (1/4,1/4) to (1/2,0) seems to have two possible explanations. Either further-NN superexchange in La$_2$NiO$_4$ yields an antiferromagnetic $J^{\prime}$ or the quantum renormalization of the spin-wave dispersion proposed for the $S=1/2$ NN SLAFMs \cite{Roennow2001_RMCH,DallaPiazza2014_DMCN,Christensen2007_CRMH,Tsyrulin2009_TPSX,Tsyrulin2010_TXSL,Wang2012_WLFE} is also present in $S=1$ systems. Testing the first proposal will require detailed electronic structure calculations of the next-NN superexchange in La$_2$NiO$_4$. This is beyond the scope of this present study. However, we note that $t/U$ is smaller in La$_2$NiO$_4$ and that other pathways such as Ni-O-O-Ni with direct overlap between the oxygens may be more important in La$_2$NiO$_4$ than La$_2$CuO$_4$ \cite{Annett1989_AMMS}. Thus, the $t^{\prime}$ hopping involving these oxygens could yield an antiferromagnetic $J^{\prime}$.  

A second proposal, that the downward dispersion is due to a renormalization of the spin wave energies, is supported by the similarity to $S=$1/2 SLAFMs: CFTD~\citep{Roennow2001_RMCH,DallaPiazza2014_DMCN,Christensen2007_CRMH}, Cu(pyrazine)$_2$(ClO$_4$)$_2$~\citep{Tsyrulin2009_TPSX,Tsyrulin2010_TXSL} and CuF$_2$(H$_2$O)$_2$(pyrazine)~\citep{Wang2012_WLFE}. Furthermore, it resembles the dispersion predicted by various theoretical models for $S=$1/2 NN SLAFMs. These models suggest that quantum effects lead to an anomaly in the dispersion at (1/2,0). The techniques used in the models are series expansions (SE) of the NN Heisenberg-Ising model~\citep{Singh1995_SiGe,Zheng2005_ZhOH}, quantum Monte-Carlo  simulations~\citep{Sandvik2001_SaSi,Shao2017_SQCC}, exact diagonalization~\citep{Luescher2009_LuLa}, continuous similarity transformations~\citep{Powalski2018_PoSU,Powalski2015_PoUS} and density matrix renormalization group (DMRG) simulations~\citep{Verresen2018_VePM}.

In CFTD~\citep{Roennow2001_RMCH,DallaPiazza2014_DMCN,Christensen2007_CRMH} and  LCO~\citep{Headings2010_HHCP}, the one-magnon spectral weight is suppressed near (1/2,0) relative to LSWT and a strong continuum with longitudinal and transverse character is observed.  LSWT predicts the same one-magnon spectral weight along the entire mBZ boundary. However, some of the other models for $S=$1/2 quantum SLAFM mentioned above predict the suppressed spectral                                                                                                                                                                                                                                                                                                                                                                                                                                                                                                                                                                                                                                                                                                                                                                                                                                                                                                                                                                                                                                                                                                                                                                                                                                                                                                                                                                                                                                                                                                                                                                                                                                                                                                                                                                                                                                                                                                                                                                                                                                                                                                                                                                                                                                                                                                                                                                                                                                                                                                                                                                                                                                                                                                                                                                                                                                                                                                                                                                                                                                                                                                                                                                                                                                                                                                                                                                                                                                                                                                                                         weight and continuum near $(1/2,0)$. We find no evidence for a wave-vector-dependent variation of the one-magnon spectral weight or continuum at (1/2,0) in LNO. 

\section{Conclusion}
Our results show that magnetic excitations in La$_2$NiO$_4$ are not described by a simple classical ($S\rightarrow\infty$) Heisenberg model with only nearest-neighbor interactions.  The energy of the spin waves disperses along the antiferromagnetic Brillouin zone boundary from (1/4,1/4) to a minimum at (1/2,0). This is in the opposite sense to that in the $S=1/2$ system La$_2$CuO$_4$, but the same sense as in other $S=1/2$ systems with smaller $t/U$ and the isostructural $S=3/2$ compound La$_2$CoO$_4$. The origin of the dispersion in La$_2$NiO$_4$ is unclear. It may be due to a quantum renormalization of the spin-wave energies or an antiferromagnetic second-nearest-neighbor superexchange. The overall intensity of the spin wave excitations is suppressed relative to linear spin-wave theory, probably due to covalency effects.  

\section{Acknowledgments}
The authors would like to thank Ruben Verresen, Roderich Moessner and James Annett for useful discussions. A.N.P. and S.M.H acknowledge funding and support from the Engineering and Physical Sciences Research Council (EPSRC) under  Grant Nos. EP/L015544/1 and EP/R011141/1.  Beamtime at ISIS and SNS were provided under proposals RB920380~\cite{Boothroyd2009_BoHa} and 26529.1 respectively. A portion of this research used resources at the Spallation Neutron Source, a DOE Office of Science User Facility operated by the Oak Ridge National Laboratory.  A.N.P. acknowledges support from the U.S. Department of Energy, Office of Science, Basic Energy Sciences, Materials Sciences and Engineering Division, under Contract No. DE-AC02-76SF00515. R.C. acknowledges support from the European Research Council under the European Union's Horizon 2020 Research and Innovation Programme Grant Agreement No.: 788814 (EQFT).

\section{Appendix}
\subsection{Neutron Scattering and Dynamic Correlation Functions}
Inelastic neutron scattering measures dynamic spin-spin correlation functions defined as
\begin{align}
S^{\alpha\beta}(\mathbf{q},\omega)= \frac{1}{2 \pi \hslash } \int_{-\infty}^{\infty}e^{-i\omega t} \langle S_{\mathbf{q}}^{\alpha}(0) S_{\mathbf{-q}}^{\beta}(t)
\rangle\, \text{d}t,
\end{align}
where $\mathbf{S}_{\mathbf{q}}$ is the Fourier transformed spin operator \cite{Marshall1971_MaLo,Boothroyd2020_Boot} (Multiply by $g^2 \mu_B^2$ to get in units of $\mu^2_B$).  Our model spin-wave calculations compute $S^{\alpha\beta}(\mathbf{q},\omega)$ which is diagonal in our case. 
In the dipole approximation the magnetic inelastic neutron scattering cross section is given by
\begin{align}\label{eqn:pardiffcross}
\dfrac{\text{d}^2\sigma}{\text{d}\Omega\text{d} E'}=&\ \dfrac{k_f}{k_i}S(\mathbf{q},\omega)\nonumber\\
=&\ \dfrac{k_f}{k_i}\left(\dfrac{\gamma r_e}{2\mu_B}\right)^2|F(\mathbf{q})|^2e^{-2W} \nonumber \\ 
&\ \ \times \sum_{\alpha \beta} \left(\delta_{\alpha\beta}-\frac{q_{\alpha} q_{\beta}}{q^2} \right) g^2\mu_B^2\,S^{\alpha\beta}(\mathbf{q},\omega),
\end{align}
where $g\approx2$, $e^{-2W}\approx 1$ for low temperatures, $F(\mathbf{q})$ is the magnetic form factor, $\mu_B$ is the Bohr magneton, and 
$(\gamma r_e/2)^2$=72.4~mb.

While the model is derived for the basal NiO$_2$ layer in Fig.~\ref{fig:La2NiO4_structure} with the spins aligned along the HTT $\mathbf{a}$-axis, $S^{aa}$ and $S^{bb}$ for neighboring NiO$_2$ planes in the LTT structure are related by a $90^\circ$ rotation around $\mathbf{c}$. This implies that in successive planes the fluctuations along the HTT $\mathbf{a}$- and $\mathbf{b}$-axes are interchanged. In the analysis this is considered by averaging the polarization factor over both axes.

Furthermore, although $S(\mathbf{q},\omega)$ is averaged over a range in $L$ by utilizing the HORACE~\cite{Ewings2016_EBLD} package the $\mathbf{q}$-vectors of the averaged pixels (neutrons) are retained and are then included in the calculations of Eqn.~\ref{eqn:pardiffcross}. It is then averaged over the calculated values of $S(\mathbf{q},\omega)$. This accounts for the $\mathbf{q}$ dependence of the polarization factor and the magnetic form factor. For more details see Ref.~\cite{Ewings2016_EBLD} and HORACE documentation.

\subsection{Derivation of Correlation Functions using Rotating Reference Frame}
\label{sec:dev}
Here we summarize the derivations of the dynamical correlations (included two-magnon correlations) within linear spin-wave theory for a square-lattice spin Hamiltonian appropriate for the $S$=1 Ni$^{2+}$ layers in La$_2$NiO$_4$ as given in Eqn.~\ref{eqn:Ham} with a collinear two-sublattice N\'{e}el magnetic structure. To the best of our knowledge the formulas have not been reported before for the Hamiltonian Eqn.~\ref{eqn:Ham} with the anisotropy terms included here.

It is convenient to transform the Hamiltonian to a rotating reference frame \cite{Coldea2003_CoTT} where the local $xyz$ spin axes at every site are defined such that $\mathbf{z}$ is along the local ordered spin direction. In this frame the ground state is ferromagnetic and the magnetic unit cell is the same as the structural primitive unit cell (one spin per cell). This can be achieved by labelling the spin axes on the N\'{e}el A sublattice (which contains the origin) with ordered spins along the $+a$ axis as ($x,y,z$) along ($b,c,a$) and for the B sublattice with spins along $-a$ as ($x,y,z$) along ($b,-c,-a$), so at a general site $\mathbf{r}$ the spin components are given by $S^x_{\mathbf{r}}=S^b_{\mathbf{r}}$, $S^y_{\mathbf{r}}=-S^a_{\mathbf{r}}\sin(\boldsymbol{\tau}_{\textrm{AF}}\cdot\mathbf{r})+S^c_{\mathbf{r}}\cos(\boldsymbol{\tau}_{\textrm{AF}}\cdot\mathbf{r})$ and $S^z_{\mathbf{r}}=S^a_{\mathbf{r}}\cos(\boldsymbol{\tau}_{\textrm{AF}}\cdot\mathbf{r})+S^c_{\mathbf{r}}\sin(\boldsymbol{\tau}_{\textrm{AF}}\cdot\mathbf{r})$ with $\boldsymbol{\tau}_{\textrm{AF}}=(1/2,1/2)$ the N\'{e}el magnetic structure propagation vector, expressed in reciprocal lattice units of the structural HTT cell. Here ($S_{\mathbf{r}}^a,S_{\mathbf{r}}^b,S_{\mathbf{r}}^c$) are the spin components along the crystallographic HTT axes. In the rotating frame the spin-wave Hamiltonian to quadratic order is obtained as,
\begin{equation}
\label{eqn:Hrot}
{\cal H}_{\rm rot}=\frac{1}{2}\sum_{\mathbf{q}} 
\left[\begin{array}{cc}
a^{\dag}_{\mathbf{q}} & a_{-\mathbf{q}}
\end{array}\right]
\left[\begin{array}{cc}
A_{\mathbf{q}} & B_{\mathbf{q}} \\
B_{\mathbf{q}} & A_{\mathbf{q}}
\end{array}\right]
\left[\begin{array}{l}
a_{\mathbf{q}}\\
a_{-\mathbf{q}}^{\dag}
\end{array}\right],
\end{equation}   
where the sum extends over all wave vectors $\mathbf{q}$ in the full (structural) Brillouin zone and $a_{\mathbf{q}}^{\dag}$ is the Fourier transformed spin creation operator. Here (including the cyclic exchange) we have,
\begin{align}
A_{\mathbf{q}}&=4S\bigg[-\frac{K_a}{2}+\frac{K_c}{4}+(J-2S^2J^\Box)   
 \nonumber \\ 
 &-(J^{\prime}-S^2J^\Box)(1-\nu_h\nu_k)-J^{\prime\prime}(2-\nu_h^2-\nu_k^2)\bigg] \nonumber \\ 
B_{\mathbf{q}}&=4S\left[(J-2S^2J^\Box)\left(\frac{\nu_h+\nu_k}{2} \right)-\frac{K_c}{4}\right],
\label{AB}
\end{align}
with $\nu_\xi=\cos(2\pi\xi)$. The $2 \times 2$ Hamiltonian matrix in Eqn.~\ref{eqn:Hrot} can be brought to diagonal form using a Bogoliubov basis transformation,
\begin{equation}
\left[\begin{array}{l}
a_{\mathbf{q}}\\
a_{-\mathbf{q}}^{\dag}
\end{array}\right]=
\left[\begin{array}{rr}
u_{\mathbf{q}} & -v_{\mathbf{q}} \\
-v_{\mathbf{q}} & u_{\mathbf{q}}
\end{array}\right]
\left[\begin{array}{l}
\alpha_{\mathbf{q}}\\
\alpha_{-\mathbf{q}}^{\dag}
\end{array}\right],
\end{equation}
where $\alpha_{\mathbf{q}}^{\dag}$ creates a spin wave with dispersion $\hslash\omega_{\mathbf{q}}=Z_c\sqrt{A_{\mathbf{q}}^2-B_{\mathbf{q}}^2}$, $u_{\mathbf{q}}=\cosh \theta_{\mathbf{q}}$, $v_{\mathbf{q}}=\sinh\theta_{\mathbf{q}}$, and $\tanh (2\theta_{\mathbf{q}})=B_{\mathbf{q}}/A_{\mathbf{q}}$. Here $Z_c$ is a dispersion renormalization factor due to higher order effects neglected at linear order in spin-wave theory. The dynamical correlations in the rotating frame (at zero temperature) are obtained as
\begin{align}
S^{xx}(\mathbf{q},\omega) &=
Z_d\frac{S}{2}\left|u_{\mathbf{q}}-v_{\mathbf{q}}\right|^2 \delta(\hslash\omega-\hslash\omega_{\mathbf{q}}) \nonumber \\
&=Z_d\frac{S}{2}\sqrt{\frac{A_{\mathbf{q}}-B_{\mathbf{q}}}{A_{\mathbf{q}}+B_{\mathbf{q}}}}\delta(\hslash\omega-\hslash\omega_{\mathbf{q}}) \\
S^{yy}(\mathbf{q},\omega)&=Z_d\frac{S}{2}\left|u_{\mathbf{q}}+v_{\mathbf{q}}\right|^2 \delta(\hslash\omega-\hslash\omega_{\mathbf{q}}) \nonumber \\
&=Z_d\frac{S}{2} \sqrt{\frac{A_{\mathbf{q}}+B_{\mathbf{q}}}{A_{\mathbf{q}}-B_{\mathbf{q}}}} \delta(\hslash\omega-\hslash\omega_{\mathbf{q}})\\
S^{zz}(\mathbf{q},\omega)&= N(S-\Delta S)^2\delta(\hslash\omega)\delta(\mathbf{q}-\bm{\tau})+ \nonumber \\
&\ \ \ \ \frac{Z_{2\rm{M}}}{2N}\sum_{\mathbf{q}_1,\mathbf{q}_2} f(\mathbf{q}_1,\mathbf{q}_2)\delta(\hslash\omega-\hslash\omega_{\mathbf{q}_1}-\hslash\omega_{\mathbf{q}_2})   \nonumber \\ &\ \ \ \ \ \ \ \ \ \ \ \ \ \ \ \ \ \times\delta(\mathbf{q}-\mathbf{q}_1-\mathbf{q}_2-\bm{\tau}).
\end{align}

The dynamical correlations in the original (fixed) reference frame (Eqns.~\ref{eqn:S_bb}-\ref{eqn:S_aa}) are obtained through Fourier transformation, such that $S^{bb}\equiv S^{xx}$ whereas $S^{aa}(\mathbf{q},\omega)=S^{zz}(\mathbf{q}+\boldsymbol{\tau}_{\textrm{AF}},\omega)$ and $S^{cc}(\mathbf{q},\omega)=S^{yy}(\mathbf{q}+\boldsymbol{\tau}_{\textrm{AF}},\omega)$, i.e. the latter two correlation functions are momentum shifted by $\boldsymbol{\tau}_{\textrm{AF}}$. 
In obtaining Eqns.~\ref{eqn:S_bb}-\ref{eqn:S_aa} we have used the fact that $2\boldsymbol{\tau}_{\textrm{AF}}$ is a vector of the reciprocal lattice of the HTT structural cell, so wavevectors $\mathbf{q}-\boldsymbol{\tau}_{\textrm{AF}}$ and $\mathbf{q}+\boldsymbol{\tau}_{\textrm{AF}}$ are equivalent by reciprocal space translational symmetry.

The in-plane (along $\mathbf{b}$) spin correlations shows a magnon mode with dispersion $\omega_{\mathbf{q}}$ (red line in Fig.~\ref{fig:Fits}(a)) with the gap 
\begin{align}
\Delta_1&=4Z_cS\sqrt{\left(2J-\frac{K_a}{2}+\frac{K_c}{2}\right)\frac{-K_a}{2}}
\nonumber \\
&\approx4Z_cS\sqrt{J(-K_a)}
\end{align}
and strong intensity above the antiferromagnetic Bragg peaks at $\bm{\tau}+\boldsymbol{\tau}_{\textrm{AF}}$, and the larger gap 
\begin{align}
\Delta_2&=4S\sqrt{\left(2J-\frac{K_a}{2}\right)\left(\frac{-K_a}{2}+\frac{K_c}{2}\right)} \nonumber \\
&\approx4Z_cS\sqrt{J(K_c-K_a)}
\end{align}
and weak intensity at $\boldsymbol{\tau}$. The out-of-plane correlations (along $\mathbf{c}$) will show the wavevector-shifted dispersion $\omega^\prime_{\mathbf{q}}=\omega_{\mathbf{q}+\boldsymbol{\tau}_{\textrm{AF}}}$ (blue line in Fig.~\ref{fig:Fits}(a)) with reversed gaps compared to $\omega_{\mathbf{q}}$, i.e. gap $\Delta_1$ at $\boldsymbol{\tau}$ and $\Delta_2$ at $\bm{\tau}+\boldsymbol{\tau}_{\textrm{AF}}$. The longitudinal correlations (along $a$) will show the elastic magnetic Bragg peaks at $\bm{\tau}+\boldsymbol{\tau}_{\textrm{AF}}$ and a two-magnon continuum, with a gap of $\Delta_1+\Delta_2$ at $\boldsymbol{\tau}$ and onsets with gaps at $2\Delta_1$ and $2\Delta_2$ at $\bm{\tau}+\boldsymbol{\tau}_{\textrm{AF}}$.

\subsection{Derivation of Correlation Functions using Antiferromagnetic Unit Cell}
\label{sec:alt_dev}
The dynamical correlation functions, derived in Appendix~\ref{sec:dev}, can also be derived in the antiferromagnetic unit cell. Utilizing this unit cell implies a doubling of the number of Ni$^{2+}$ ions per unit cell and thus, an effective doubling of the spin-wave modes but does not require the transformation of the Hamiltonian to a rotating reference frame. The notations used in this section are the same as in Appendix~\ref{sec:dev}.

For the antiferromagnetic unit cell the spin-wave Hamiltonian to quadratic order can be written as
\begin{align}
{\cal H}\!=\!\!\frac{1}{2}\!\!\sum_{\mathbf{q}}\!\!
\left[\!\!\begin{array}{c}
a^{\dag}_{\mathbf{q}} \\ b_{\mathbf{q}} \\ a_{\mathbf{-q}} \\ b^{\dag}_{\mathbf{-q}}
\!\end{array}\!\!\right]^{\!\!T}\!\!
\left[\!\!\begin{array}{cccc}
A_{\mathbf{q}} & B^{\prime}_{\mathbf{q}} & -SK_c & 0\\
B^{\prime}_{\mathbf{q}} & A_{\mathbf{q}} & 0 & -SK_c\\
-SK_c & 0 & A_{\mathbf{q}} & B^{\prime}_{\mathbf{q}}\\
0 & -SK_c & B^{\prime}_{\mathbf{q}} & A_{\mathbf{q}} \\
\end{array}\!\!\right]\!\!\!
\left[\!\!\begin{array}{c}
a_{\mathbf{q}} \\ b^{\dag}_{\mathbf{q}} \\ a^{\dag}_{-\mathbf{q}} \\ b_{-\mathbf{q}}
\end{array}\!\!\right],
\end{align}
where $B^{\prime}_{\mathbf{q}}=B_{\mathbf{q}}+SK_c$, $b^{\dag}$ is the local spin deviation creation operator on the B sublattice, the sum extends over all wave vectors in the mBZ and $A_{\mathbf{q}}$, $B_{\mathbf{q}}$ are the same as in Eqn.~\ref{eq:DispersionLNO}. As can be seen, there are two flavors of operators.
This 4 × 4 Hamiltonian matrix again can be brought to diagonal form using the following Bogoliubov basis transformation including two flavors of new spin wave creation (annihilation) operators, corresponding to spin-wave modes polarized along $\mathbf{b}$ and $\mathbf{c}$,
\begin{equation}
\left[\!\!\begin{array}{l}
a_{\mathbf{q}}\\
b_{\mathbf{q}}^{\dag}\\
a_{-\mathbf{q}}^{\dag}\\
b_{-\mathbf{q}}
\end{array}\!\!\right]=\frac{1}{\sqrt{2}}\!\!
\left[\begin{array}{rrrr}
u_{\mathbf{q}} & -v^\prime_{\mathbf{q}} & -v_{\mathbf{q}} & -u^\prime_{\mathbf{q}} \\
-v_{\mathbf{q}} & u^\prime_{\mathbf{q}} & u_{\mathbf{q}} & v^\prime_{\mathbf{q}} \\
-v_{\mathbf{q}} & -u^\prime_{\mathbf{q}} & u_{\mathbf{q}} & -v^\prime_{\mathbf{q}} \\
u_{\mathbf{q}} & v^\prime_{\mathbf{q}} & -v_{\mathbf{q}} & u^\prime_{\mathbf{q}} 
\end{array}\right]
\left[\!\!\begin{array}{l}
\alpha_{\mathbf{q}}\\
\beta_{\mathbf{q}}^{\dag}\\
\alpha_{-\mathbf{q}}^{\dag}\\
\beta_{-\mathbf{q}}
\end{array}\!\!\right],
\end{equation}

The transformation thus yields two sets of terms, $u_{\mathbf{q}}$, $v_{\mathbf{q}}$ and $\omega_{\mathbf{q}}$ and $u_{\mathbf{q}}^\prime=\cosh\theta_{\mathbf{q}}^\prime$,   $v_{\mathbf{q}}^\prime=\sinh\theta_{\mathbf{q}}^\prime$ and $\omega_{\mathbf{q}}^\prime$ with,
\begin{align}
\tanh (2\theta_{\mathbf{q}})&=\frac{B^{\prime}_{\mathbf{q}}-SK_c}{A_{\mathbf{q}}}=\frac{B_{\mathbf{q}}}{A_{\mathbf{q}}},\\
\tanh (2\theta^\prime_{\mathbf{q}})&=\frac{B^{\prime}_{\mathbf{q}}+SK_c}{A_{\mathbf{q}}}\nonumber\\
&=-\frac{B_{\mathbf{q}+\bm{\tau}_{\mathrm{AF}}}}{A_{\mathbf{q}+\bm{\tau}_{\mathrm{AF}}}}\ \ =-\tanh (2\theta_{\mathbf{q}+\bm{\tau}_{\mathrm{AF}}}),\nonumber
\end{align}

 and
 
\begin{align}
\hslash\omega_{\mathbf{q}}&=Z_c\sqrt{A_{\mathbf{q}}^2-(B^\prime_{\mathbf{q}}-SK_c)^2}=Z_c\sqrt{A_{\mathbf{q}}^2-B_{\mathbf{q}}^2},\\
\hslash\omega_{\mathbf{q}}^\prime&=Z_c\sqrt{A_{\mathbf{q}}^2-(B^\prime_{\mathbf{q}}+SK_c)^2}\nonumber\\
&=Z_c\sqrt{A_{\mathbf{q+\bm{\tau}_{\mathrm{AF}}}}^2-B_{\mathbf{q}+\bm{\tau}_{\mathrm{AF}}}^2}\ \ \ \,=\hslash\omega_{\mathbf{q}+\bm{\tau}_{\mathrm{AF}}}. \nonumber
\end{align}

Here we again use the fact that $2\boldsymbol{\tau}_{\textrm{AF}}$ is a reciprocal lattice vector.

The correlations functions $S^{bb}$ and $S^{cc}$ follow as,
\begin{align}
S^{bb}(\mathbf{q},\omega)\!=&\frac{Z_dS}{2}\left|u_{\mathbf{q}}\!-\!v_{\mathbf{q}}\right|^2 \delta(\hslash\omega\!-\!\hslash\omega_{\mathbf{q}})\label{eqn:bb_alt}\\
S^{cc}(\mathbf{q},\omega)\!=&\frac{Z_dS}{2}\left|u^\prime_{\mathbf{q}}\!-\!v^\prime_{\mathbf{q}}\right|^2 \delta(\hslash\omega\!-\!\hslash\omega_{\mathbf{q}}^\prime)\label{eqn:cc_alt}\\
=&\frac{Z_dS}{2}\left|u_{\mathbf{q}+\bm{\tau}_{\mathrm{AF}}}\!+\!v_{\mathbf{q}+\bm{\tau}_{\mathrm{AF}}}\right|^2 \delta(\hslash\omega\!+\!\hslash\omega_{\mathbf{q}+\bm{\tau}_{\mathrm{AF}}})\nonumber.
\end{align}
Using the relations between $\tanh (2\theta_{\mathbf{q}})$ and $\tanh (2\theta^\prime_{\mathbf{q}})$ through the translation by $\bm{\tau}_{\mathrm{AF}}$ from Appendix~\ref{sec:dev}, the Eqns.~\ref{eqn:bb_alt}-\ref{eqn:cc_alt} can be written as Eqns.~\ref{eqn:S_bb}-\ref{eqn:S_cc} and thus, yield the same results as the rotating frame method.

The longitudinal dynamical correlations take the form,
\begin{align}
S^{aa}(\mathbf{q},\omega)& =N(S-\Delta S)^2\delta(\hslash\omega)\delta(\mathbf{q}-\bm{\tau}-\bm{\tau}_{\mathrm{AF}})+ \nonumber \\
\frac{Z_{2\rm{M}}}{2N}&\sum_{\mathbf{q}_1,\mathbf{q}_2} f^\prime(\mathbf{q}_1,\mathbf{q}_2)\delta(\hslash\omega-\hslash\omega_{\mathbf{q}_1}-\hslash\omega_{\mathbf{q}_2}^\prime)  \times \nonumber \\ &\delta(\mathbf{q}-\mathbf{q}_1-\mathbf{q}_2-\bm{\tau}), \label{eqn:aa_alt}
\end{align}
where $f^\prime(\mathbf{q}_1,\mathbf{q}_2)=\left|v_{\mathbf{q}_1}u^\prime_{\mathbf{q}_2}-u_{\mathbf{q}_1}v^\prime_{\mathbf{q}_2}\right|^2$. Using the transformations relations, we find that,
\begin{align}
f^\prime(\mathbf{q}_1,\mathbf{q}_2)&=\left|v_{\mathbf{q}_1}u^\prime_{\mathbf{q}_2}-u_{\mathbf{q}_1}v^\prime_{\mathbf{q}_2} \right|^2\nonumber\\
&=\left|v_{\mathbf{q}_1}u_{\mathbf{q}_2+\boldsymbol{\tau}_{\textrm{AF}}}+u_{\mathbf{q}_1}v_{\mathbf{q}_2+\boldsymbol{\tau}_{\textrm{AF}}}\right|^2\nonumber\\
&=f(\mathbf{q}_1,\mathbf{q}_2+\boldsymbol{\tau}_{\textrm{AF}}),
\end{align}
Thus, we can transform Eqn.~\ref{eqn:aa_alt} to yield Eqn.~\ref{eqn:S_aa}. In the antiferromagnetic unit cell description the two spin-wave modes appear mixed in the two-magnon scattering and applying the shift by $\boldsymbol{\tau}_{\textrm{AF}}$ effectively `decouples' the modes.

\subsection{Dzyaloshinskii–Moriya interaction}
\label{sec:dmi}
A finite Dzyaloshinskii–Moriya interaction (DMI) in a N\'eel SLAFM yield two non-degenerate spin-wave modes similar to the hard-axis anisotropy $K_c$. Contrary, to a hard-axis anisotropy DMI also yields a spin canting which can be estimated from Eq.~3 in Ref.~\cite{Yamada1992_YONH}. The reported spin canting of 0.1$^\circ$ implies a DMI of $\leq0.1$\,meV. Conversely, to establish the observed gap between the two spin-wave modes~\cite{Nakajima1993_NYHO} for $K_c=0$ numeric LSWT calculations~\cite{Toth2012_TLMP} suggest a required DMI of $\approx3.5$\,meV, yielding a spin canting of $\approx3.5^\circ$, which is $35\times$ larger than the reported value~\cite{Yamada1992_YONH}. So, the effect from the DMI on the dynamics is much smaller than the effect from the $K_c$ and is hence negligible.
\vfill

%

\end{document}